\documentclass[a4paper,11pt]{article}

\pdfoutput=1

\usepackage{jheppub,amsmath,subfig}
\definecolor{green}{rgb}{0.1,0.8,0.2}

\title{6d Dual Conformal Symmetry and Minimal Volumes in AdS}
 
\author{Jyotirmoy Bhattacharya and Arthur E. Lipstein }

\affiliation{Centre for Particle Theory \& Department of Mathematical Sciences, Durham University, South Road, Durham DH1 3LE, United Kingdom.}

\emailAdd{jyotirmoy.bhattacharya@durham.ac.uk, arthur.lipstein@durham.ac.uk}

\abstract{The S-matrix of a theory often exhibits symmetries which are not manifest
from the viewpoint of its Lagrangian. For instance, powerful constraints 
on scattering amplitudes are imposed by the dual conformal symmetry of 
planar 4d $\mathcal{N}=4$ super Yang-Mills theory and the ABJM theory.
Motivated by this, we investigate the consequences of dual conformal symmetry in six dimensions,
which may provide useful insight into the worldvolume theory of M5-branes (if it enjoys such a symmetry). 
We find that 6d dual conformal symmetry uniquely 
fixes the integrand of the one-loop 4-point amplitude, and its structure suggests a Lagrangian with more than two derivatives. 
On integrating out the loop momentum in $6-2 \epsilon$ dimensions, the result is very similar
to the corresponding amplitude of $\mathcal{N}=4$ super Yang-Mills theory. We confirm this result holographically by generalizing the Alday-Maldacena solution for a minimal area string in Anti-de Sitter space to a minimal volume M2-brane ending on a pillow-shaped surface in the boundary whose seams correspond to a null-polygon. This involves careful 
treatment of a prefactor which diverges as $1/\epsilon$, and we comment on its possible interpretation. 
We also study 2-loop 4-point integrands with 6d dual conformal symmetry and speculate on the existence of an all-loop formula for the 4-point amplitude.
}

\begin{document}

% %preprint
% \begin{flushright} 
% % \small{preprint} 
% \end{flushright}

%
\maketitle
\flushbottom
%
%%%%%%%%%%%%%%%%%%%%%%%%%%%%%

%****************************************
\section{Introduction}
%****************************************

The AdS/CFT conjecture \cite{Maldacena:1997re} establishes a correspondence between 
string or M-theory in the near horizon geometry of a stack of 
D-branes or M-branes to the effective low-energy world-volume theory of the branes.
Although this idea has now grown into a wide framework of gauge-gravity duality with 
diverse applications, the most prominent and early examples of this conjecture involve maximally 
supersymmetric field theories on D3, M2 and M5-branes, which are conjectured to be dual to 
IIB string theory on AdS$_5 \times$S$^5$ \cite{Green:1981yb}, M-theory on AdS$_4 \times$S$^7$ \cite{Freund:1980xh},
and AdS$_7 \times$S$^4$ \cite{Pilch:1984xy}, respectively. 

In the first example, the worldvolume theory on D3-branes
is 4d $\mathcal{N}=4$ super Yang-Mills theory (SYM) \cite{Brink:1976bc}, which admits an expansion in terms of the 't Hooft parameter, which in turn can also be classified 
by the topologies of Riemann surfaces \cite{'tHooft:1973jz}. In such an expansion, the topologies with higher genus 
are suppressed by the rank of the gauge group $N$, and in the limit when $N$ is large, 
the theory is entirely dominated by planar diagrams. In the planar limit, the theory is 
believed to be solvable (for a review, see for example \cite{Beisert:2010jr}). 

The second and third examples of the AdS/CFT conjecture are more challenging because M-theory arises as the
strong coupling limit of string theory, making it difficult to formulate the worldvolume 
theories of M-branes. This was however, recently accomplished for M2-branes \cite{Aharony:2008ug}. 
The key insight was to perform a $\mathbb Z_k$ orbifold of the space transverse to the branes which breaks $1/4$ of 
the susy when $k>2$, but allows one to define a tunable coupling for the theory.  
The Lagrangian is a 3d superconformal Chern-Simons theory with $\mathcal{N}=6$ susy known as the ABJM theory
\footnote{Prior to the ABJM theory, there were other proposals which have maximal superconformal 
symmetry, although the interpretation of these theories is not fully understood 
\cite{Gustavsson:2007vu,Bagger:2007jr,Bagger:2007vi}.}. When the orbifold parameter is large, 
it is dual to IIA string theory on AdS$_4\times$CP$^3$. 

The third example is the least understood 
because the worldvolume theory describing multiple M5-branes remains 
elusive. The M5-brane theory is not only crucial for understanding AdS/CFT 
and M-theory, it also provides the geometric origin of electric-magnetic 
duality of many supersymmetric theories that arise from dimensional 
reduction \cite{Gaiotto:2009we,Alday:2009aq,Dimofte:2011ju}. 
It is believed to be a 6d CFT whose field content is a $(2,0)$ 
tensor multiplet, which consists of a self-dual two-form gauge field, 
five scalars, and eight fermions. Although the worldvolume theory for a 
single M5-brane is well-understood \cite{Perry:1996mk,Pasti:1997gx}, 
it is unclear how to generalize it to more than one. In the absence of a tunable coupling, 
it is not clear whether such a theory would even admit a Lagrangian description (see 
\cite{Aharony:1997th,ArkaniHamed:2001ie,Lambert:2010wm,Saemann:2012uq,Samtleben:2012fb,Douglas:2010iu,Lambert:2010iw,Lipstein:2014vca}   
for some attempts in this direction). 

Using modern methods for computing on-shell scattering amplitudes, 
it is possible to learn a great deal about the S-matrix of many theories 
without reference to their Lagrangians. Moreover, the S-matrix often exhibits 
symmetries that are totally obscure from the point of view of the Lagrangian. 
For example, the planar amplitudes of $\mathcal{N}=4$ SYM \cite{Drummond:2006rz,Brandhuber:2008pf,Drummond:2008vq} and the ABJM theory \cite{Bargheer:2010hn,Huang:2010qy,Gang:2010gy} 
exhibit a remarkable property known as dual conformal symmetry 
in the planar limit.
In the context of $\mathcal{N}=4$ SYM, the dual conformal symmetry can 
be understood as the ordinary conformal symmetry of a dual Wilson loop 
whose contour is obtained by arranging the external momenta of the amplitude head to tail  
\cite{Alday:2007hr,Brandhuber:2007yx,Drummond:2007cf,Drummond:2007aua,Mason:2010yk,CaronHuot:2010ek}. 
Moreover, the amplitude/Wilson loop duality of 4d $\mathcal{N}=4$ SYM can be 
derived from the self-duality of IIB string theory under a certain combination of 
bosonic and fermionic T-duality transformations \cite{Berkovits:2008ic,Beisert:2008iq}. In the case of ABJM theory, 
the amplitude/Wilson loop duality does not generalize beyond 4-points 
and the status of the fermionic T-duality in the gravity dual is also unclear \cite{Adam:2010hh,Bakhmatov:2010fp,Colgain:2016gdj}. 
Hence, the origin of dual conformal symmetry in the ABJM theory appears to be rather mysterious.   

Note that dual conformal symmetry is not equivalent to ordinary 
superconformal symmetry and when the two are combined, 
they give rise to infinite dimensional Yangian symmetry 
for scattering amplitudes\cite{Dolan:2004ps,Drummond:2009fd}. 
Hence, dual conformal symmetry imposes very powerful constraints on the planar S-matrix. 
For example, in $\mathcal{N}=4$ SYM and ABJM one can use dual conformal symmetry 
to uniquely fix the 1-loop 4-point integrand, 
from which the 4-point tree-level amplitudes can be deduced using unitarity methods \cite{Bern:1994zx,Bern:1994cg}. 
The rest of the tree-level S-matrix can then be deduced using BCFW recursion \cite{Britto:2005fq}, 
which in principle can be used to deduce the Lagrangian. Hence, apriori if we had no idea how to 
formulate  ABJM or $\mathcal{N}=4$ SYM, we could have deduced these theories 
simply by looking for dual conformal invariant S-matrices in 3d and 4d, respectively. 

At loop-level, dual conformal symmetry is broken by IR divergences but the four 
and five points amplitudes of $\mathcal{N}=4$ SYM are nevertheless fixed to all 
orders by an anomalous dual conformal Ward identity \cite{Drummond:2007au}. These all-loop formulae, 
first conjectured by Bern, Dixon, and Smirnov  \cite{Bern:2005iz}, boil down to exponentiating 
the one-loop amplitude and encoding the coupling dependence through 
the cusp and collinear anomalous dimensions. 
This has been confirmed up to four loops in perturbation theory 
\cite{Bern:2006ew,Cachazo:2006tj,Bern:2008ap}, as well as at strong coupling using 
string theory \cite{Alday:2007hr,Alday:2007he,Alday:2009dv}. There is also evidence that a similar all-loop 
formula exists for the 4-point amplitude of the ABJM theory \cite{Henn:2010ps,Chen:2011vv,Bianchi:2011dg,Bianchi:2011fc,Bianchi:2011aa,Bianchi:2013pva,Bianchi:2014iia}. 
In particular, the 2-loop 4-pt amplitude of the ABJM theory has an almost identical structure to the 1-loop 4-point 
amplitude of $\mathcal{N}=4$ SYM, and the strong coupling 
calculation in the dual string theory is identical to that of $\mathcal{N}=4$ SYM at leading order.

Given the important role played by dual conformal symmetry in the first 
two examples of the AdS/CFT duality described above, in this note we proceed to explore its implications for the theory of 
M5-branes, assuming it enjoys such a symmetry. There are of course, several subtleties with this point of view. 
For instance, it has previously been shown that is not possible to construct a tree-level S-matrix 
for $(2,0)$ tensor multiplets assuming locality and unitarity \cite{Huang:2010rn}. 
On the other hand, since the $(2,0)$ theory is strongly coupled, it is unclear what the asymptotic states should be. In this note, we will not make any assumptions about the asymptotic states, locality, or even the superconformal symmetry of the theory. Our only assumptions will be that the theory admits something analogous to a planar 
limit where a semi-classical supergravity description is admissible in the bulk, 
and that it has rational loop integrands 
\footnote{Since the M5-brane theory is self-dual, it is unclear how to define a topological expansion in 
terms of an 't Hooft parameter. A similar difficulty occurs for the M2-brane theory, but can be overcome by 
orbifolding the space transverse to the branes, which introduces a tunable coupling.} 
which exhibit dual conformal symmetry in six dimensions \footnote{Note that 6d $(1,1)$ SYM amplitudes have dual 
conformal symmetry in the four-dimensional sense \cite{Dennen:2010dh}, 
as do maximal SYM amplitudes in 10d \cite{CaronHuot:2010rj} and 3d \cite{Lipstein:2012kd}.}.

In this note, we find that 6d dual conformal symmetry fixes the 1-loop 4-point integrand. 
Like $\mathcal{N}=4$ SYM, the integrand can be associated with a scalar box diagram with massless external 
legs, but unlike $\mathcal{N}=4$ SYM, some of the propagators are squared which suggests  an underlying  Lagrangian with more than two derivative. 
We also consider triangle and bubble diagrams with dual conformal integrands 
but show that they are not important at 4 points. 
Remarkably, although the 4-point 1-loop integrand that we obtain 
has a very different spacetime structure than that of $\mathcal{N}=4$ SYM, 
its integral is essentially the same (up to scheme-dependent terms). 
Recalling that a very similar structure was found for the 2-loop 4-point amplitude of the ABJM theory, 
it therefore appears that 4-point amplitudes with dual conformal symmetry 
have a universal structure. We also carry out a preliminary analysis of 2-loop 4-point integrands consistent with 6d dual conformal symmetry 
and unitarity. Remarkably, although these integrands have a very different spacetime structure 
than that of $\mathcal{N}=4$ SYM, their Mellin-Barnes representation 
is very similar, suggesting that the similarity to $\mathcal{N}=4$ SYM might extend beyond 1-loop.

The all-loop formula for the four-point amplitude of $\mathcal{N}=4$ 
SYM was first confirmed at strong coupling by Alday and Maldacena who computed 
the minimal area of a string in AdS whose boundary corresponds to the null polygon obtained 
by arranging the external momenta of the amplitude head to tail 
 \cite{Alday:2007hr}. In fact, this was the first hint of the amplitude/Wilson loop duality. 
For the M5-brane theory, we would expect the analogous calculation to involve 
the minimal volume of an M2-brane whose boundary is a 2d surface which somehow 
encodes a null polygon. Remarkably, by relaxing a constraint of the Alday-Maldacena 
solution, we obtain a solution which minimizes the volume  of a 2-brane 
in AdS and the boundary of this solution is a pillow-shaped 
surface whose seams correspond to a null polygon
\footnote{Note that since the original null polygon of \cite{Alday:2007hr} is embedded within this more general surface, it is 
parameterized by the same two parameters corresponding to the Mandelstam variables $s$ and $t$ in 
the amplitude picture.}.  We evaluate the on-shell action for the 2-brane solution and find a structure very similar to the 6d 1-loop amplitude, which suggests an amplitude/Wilson surface duality and the existence of an all-loop BDS-like formula for the 4-point amplitude. Unlike the on-shell action of the Alday-Maldacena string solution, the on-shell action of the 2-brane has a prefactor that depends on the dimensional regularization parameter $\epsilon = (6-d)/2$ and diverges like $1/\epsilon$  as $\epsilon\rightarrow 0$. This  additional divergence is consistent with  previous 
holographic calculations of Wilson surfaces \cite{Berenstein:1998ij,Young:2011aa}, and can be associated with the 
conformal anomaly of Wilson surfaces \cite{Graham:1999pm,Henningson:1999xi,Gustavsson:2003hn,Gustavsson:2004gj}
\footnote{Also see \cite{Mori:2014tca,Corrado:1999pi} for related discussion of Wilson surface operators in the M5-brane theory.}. 

This note is organized as follows. In \ref{sec:6dAmp} 
we explore the consequences of dual conformal symmetry in six dimensions. 
In particular, we find that dual conformal symmetry fixes the 
integrand of the 1-loop 4-point amplitude, which we integrate in $d=6-2 \epsilon$ 
dimensions using the Mellin-Barnes technique. We also initiate the study of 2-loop 4-point amplitudes. 
In \S \ref{sec:minvol}, we find a solution describing a minimal volume 2-brane in AdS whose boundary is a 2d surface encoding a null polygon. 
We then compute the on-shell action for this 
solution and find that it has a very similar structure to the 1-loop 
4-point amplitude we computed in \S \ref{sec:6dAmp}, 
suggesting the existence of an amplitude/Wilson surface duality and an all-loop formula for the 4-point amplitude analogous to the BDS 
formula of $\mathcal{N}=4$ SYM. 
Finally, in \S \ref{sec:disco} we present our conclusions and describe future directions. Appendix \ref{app:6d4p2l} gives more details about 2-loop 4-point integrands and Appendix \ref{app:caldet} provides more details about the calculation of the on-shell action of the 2-brane in AdS.    

%****************************************
\section{6d Dual Conformal Amplitudes}\label{sec:6dAmp}
%****************************************

Let us consider a hypothetical 6d theory for which a planar limit can be defined and whose loop integrands are rational. We shall study the consequences of dual conformal symmetry for such a theory, primarily focusing on the 
4-point case. We shall also assume that in the planar limit it is possible to define color-ordered amplitudes which are cyclically symmetric (the definition of color-ordering in gauge theories can be found in many reviews, see for example \cite{Dixon:2013uaa}). 
In $\mathcal{N}=4$ SYM and the ABJM theory, it is possible to encode the four-point amplitudes with all possible asymptotic states into a single quantity known as a superamplitude. When this quantity is divided by the tree-level four-point superamplitude, the resulting function does not depend on asymptotic states and is determined to all orders by dual conformal symmetry in the planar limit. Moreover, it can be computed at strong coupling using holographic methods. Our goal in the present paper is to deduce such a function in six dimensions, which only relies on the assumption of dual conformal symmetry and does not require knowing the asymptotic states. The four-point superamplitude can then in principle be obtained by multiplying this quantity by the tree-level four-point superamplitude, whose structure is an important open question. We shall therefore define the quantity $\mathcal{M}_{n}=\mathcal{A}_{n}/\mathcal{A}^{\rm{tree}}_{n}$. This quantity can then be expanded in powers of the 't Hooft coupling $\lambda$:
\begin{equation}
\mathcal{M}_{n}=1+\sum_{l=1}^{\infty}\lambda^{l}\mathcal{M}_{n}^{(l)}.
\label{expansion}
\end{equation}
In the next subsections we will review the concept of dual conformal symmetry, use it to deduce the 1-loop 4-point contribution $\mathcal{M}_{4}^{(1)}$, and analyze the possible structure of $\mathcal{M}_{4}^{(2)}$.  

%--------------------------------------------------
\subsection{Review of Dual Conformal Symmetry}
%--------------------------------------------------

\begin{figure}[h]
 \centering
 \includegraphics[width=0.6\textwidth]{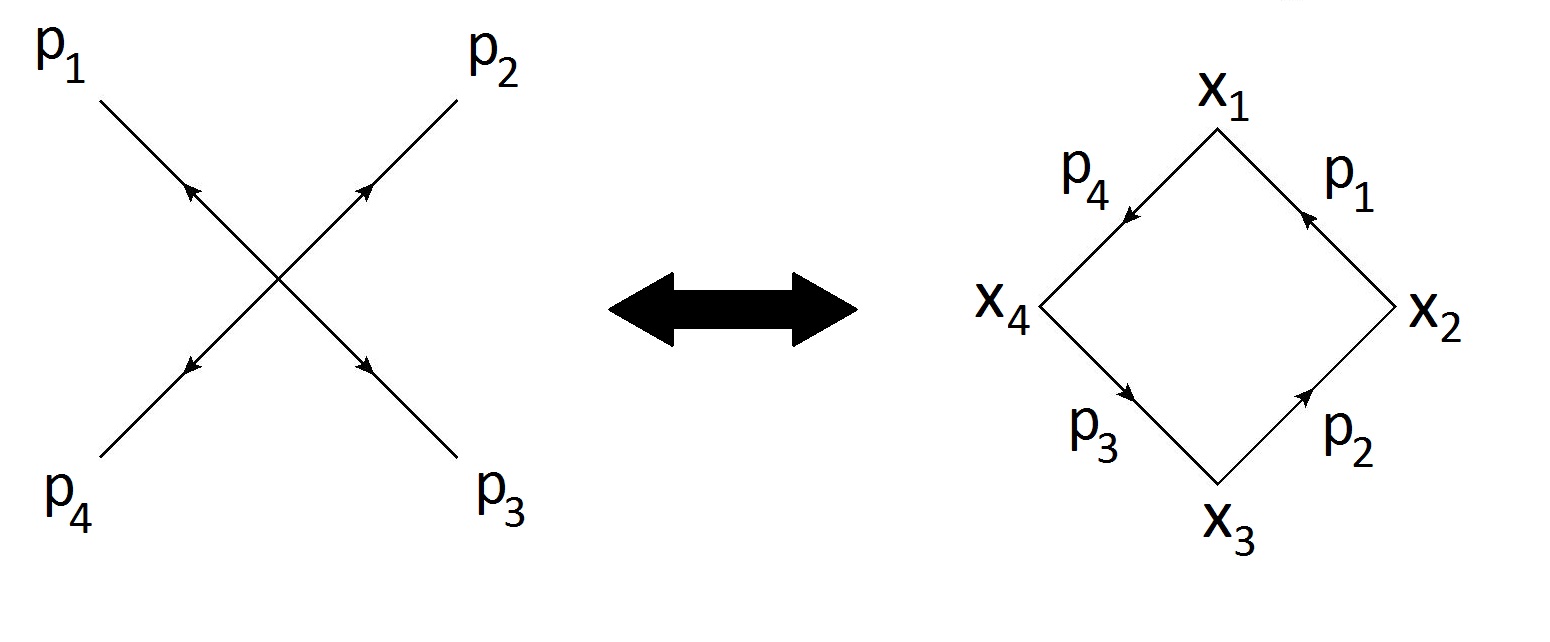}
 \caption{Arranging the external momenta of a scattering head-to-tail gives a null polygon whose vertices are coordinates of the dual space.}
 \label{ampwl}
\end{figure}

Dual conformal symmetry can be seen by arranging the external momenta of an amplitude into a polygon
and writing the amplitude as a function of the vertices of this polygon as shown in fig. \ref{ampwl}.
Note that this relies on the ability to cyclically order the external
particles and is therefore only well-defined in the planar limit.
In equations, the coordinates of the dual space
are defined by 
\[
x_{i}^\mu-x_{i+1}^\mu=p_i^\mu
\]
where $p_i^\mu$ is the momentum of the $i$th leg. These coordinates automatically incorporate momentum conservation. Dual superconformal symmetry then corresponds to conformal symmetry in the dual space. Note that the amplitudes are manifestly invariant under dual translations since they depend on external momenta which correspond to differences of points in the dual space. Dual conformal symmetry implies the nontrivial property that they transform covariantly under inversions in the dual space:
\begin{equation}
x_{i}^{\mu}\rightarrow\frac{x_{i}^{\mu}}{x_{i}^{2}}.
\label{inversion}
\end{equation}

Note that dual conformal symmetry is not equivalent to ordinary conformal symmetry and imposes very powerful constraints on scattering amplitudes. For example, in $\mathcal{N}=4$ SYM, although this symmetry is broken by IR divergences, it nevertheless determines the finite part of the four and five-point amplitudes to all orders in the t-Hooft coupling via an anomalous dual conformal Ward identy. Beyond five points it is possible to introduce a function of dual conformal cross ratios and still satisfy the anomalous dual conformal Ward identity, so the amplitudes are given by the BDS formula times a nontrivial remainder function. 

%--------------------------------------------------
\subsection{One-loop Four-point}
%--------------------------------------------------

\begin{figure}[h]
 \centering
 \includegraphics[width=0.28\textwidth]{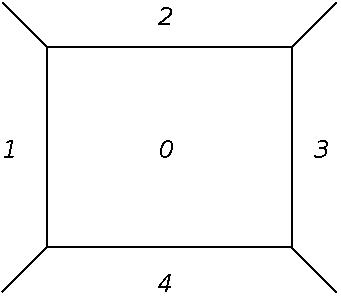}
 \caption{One loop box.}
 \label{fig:1loop}
\end{figure}

In 6d, dual conformal symmetry and cyclic invariance fix the one-loop four point integrand to have the following form:
\begin{equation}
\mathcal{M}_{4}^{(1)}=\frac{1}{8}\int\frac{d^{D}x_{0}x_{13}^{2}x_{24}^{2}}{x_{10}^{2}x_{20}^{2}x_{30}^{2}x_{40}^{2}}\left(\frac{x_{13}^{2}}{x_{01}^{2}x_{03}^{2}}+\frac{x_{24}^{2}}{x_{02}^{2}x_{04}^{2}}\right).
\label{4pt1loopintegrand}
\end{equation}
where $x_{ij}=x_i-x_j$ and the factor of $\frac{1}{8}$ is a convention tied to our definition of the coupling $\lambda$ in \eqref{expansion}. This can be represented using a 1-loop box diagram depicted in Figure \ref{fig:1loop}, where each edge of the box corresponds to a propagator in the integrand. It is also possible to define dual conformal triangle and bubble integrals, but as we describe in the next subsection, the former do not contribute at 4-points and the latter evaluate to a constant and woudl therefore only modify scheme-dependent terms in the 4-point 1-loop amplitude. Note that the integrand in \eqref{4pt1loopintegrand} is manifestly invariant under translations in the dual space. It is also easy to see that when $D=6$, it is invariant under dual inversions, under which the measure and propagators transform as follows:
\[
dx_{0}^{D}\rightarrow\frac{dx^D_{0}}{\left(x_{0}^{2}\right)^{D}},\,\,\, x_{ij}^{2}\rightarrow\frac{x_{ij}^{2}}{x_{i}^{2}x_{j}^{2}}.
\]
Since dual translations and inversions are sufficient to generate the dual conformal group, this establishes dual conformal symmetry of the integrand when $D=6$. An unusual feature of the integrand is that two of the propagators in each term are squared, which suggests that if there is an underlying Lagrangian, it may correspond to a higher derivative theory. It is worth noting that four-derivative terms appear in certain non-unitary 6d superconformal theories \cite{Ivanov:2005qf,Beccaria:2015uta}, so it would interesting to study the scattering amplitudes of these theories. Another possibility is that the underlying theory is unitary but non-local. We will return to these points in section \ref{sec:disco}. 

The integral in \eqref{4pt1loopintegrand} is IR divergent, 
so we will regulate it by taking $D=6-2 \epsilon$, where $\epsilon<0$. 
Although this breaks dual conformal symmetry, it does so in a controlled and well-understood way as we will see shortly. 

\begin{figure}[h]
 \centering
 \includegraphics[width=0.5\textwidth]{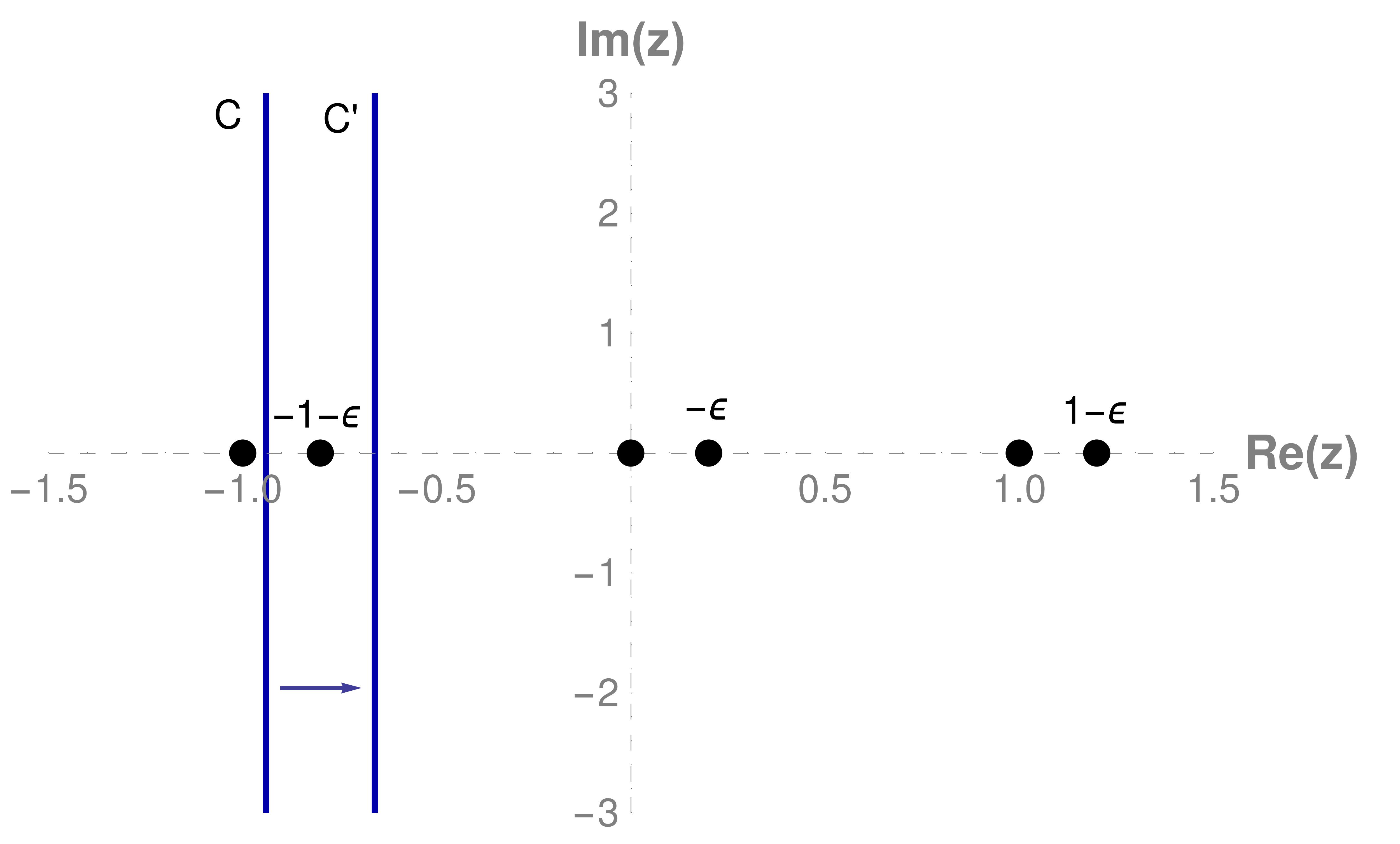}
 \caption{The contour for the integral in \eqref{4m}.}
 \label{fig:contour}
\end{figure}

This integral can then be evaluated using Mellin-Barnes techniques, as described in \cite{SmirnovBook}. In general, an integral of this form has the following Mellin-Barnes representation:
\[
\int\frac{d^{D}x_{0}}{\Pi_{i=1}^{4}\left(x_{0i}^{2}\right)^{a_{i}}}=\frac{\left(-1\right)^{a}i\pi^{D/2}}{\Gamma\left(D-a\right)\Pi_{i=1}^{4}\Gamma\left(a_{i}\right)t^{a-D/2}}\int_{c-i\infty}^{c+i\infty}\frac{dz}{2\pi i}\left(\frac{s}{t}\right)^{z}\Gamma\left(-z\right)\Gamma\left(a-D/2+z\right)
\]
\[
\times\Gamma\left(a_{1}+z\right)\Gamma\left(a_{3}+z\right)\Gamma\left(D/2-a_{1}-a_{3}-a_{4}-z\right)\Gamma\left(D/2-a_{1}-a_{2}-a_{3}-z\right),
\]
where $s=x_{13}^{2}$, $t=x_{24}^{2}$, $a=\sum_{i=1}^{4}a_{i}$, and $c$ is a complex number chosen such that 
the arguments of the gamma functions have positive real part. For the integral in \eqref{4pt1loopintegrand} with $D=6-2\epsilon$, we obtain  
\begin{equation}
\mathcal{M}_{4}^{(1)}=\frac{st^{-1-\epsilon}}{8\Gamma(-2\epsilon)}\int_{c-i\infty}^{c+i\infty}\frac{dz}{2\pi i}\left(\frac{s}{t}\right)^{z}\Gamma(-z)\Gamma(3+\epsilon+z)\Gamma(1+z)^{2}\Gamma\left(-1-\epsilon-z\right)^{2}+s\leftrightarrow t.
\label{4m}
\end{equation}
Since $\epsilon<0$, one must choose $-1<\Re(c)<-1-\epsilon$ so
that the arguments of the gamma functions have positive real part. The integral generates poles in $\epsilon$ which arise from the terms $\Gamma(1+z)^{2}\Gamma\left(-1-\epsilon-z\right)^{2}$. If we shift the contour to the right while picking up the residue at $z=-1-\epsilon$ as depicted in Figure \ref{fig:contour}, the integral over the new contour does not have poles in $\epsilon$ and is actually $\mathcal{O}(\epsilon)$ due to the prefactor of $1/\Gamma{(-2\epsilon)}$. Hence, to $\mathcal{O}(\epsilon)$ the amplitude is given by the residue at $z=-1-\epsilon$:    
\begin{equation}
e^{\epsilon\left(\gamma_E-2\right)}\mathcal{M}_{4}^{(1)}=-\frac{1}{2 \epsilon^{2}}\left(\left(\frac{\mu^{2}}{s}\right)^{\epsilon}+\left(\frac{\mu^{2}}{t}\right)^{\epsilon}\right)+\frac{1}{4}\ln^{2}\frac{s}{t}+\frac{\pi^{2}}{3}+\frac{1}{8}
\label{1loop4pt}
\end{equation}
where $\gamma_E$ is the Euler-Mascheroni constant and $\mu^2$ is the renormalization scale.

Note that our result for the 1-loop 4-point amplitude six dimensions is the same as the 1-loop amplitude of $\mathcal{N}=4$ SYM, except for the constant term. Had we multiplied by $e^{\epsilon \gamma}$ than $e^{\epsilon\left(\gamma_E-2\right)}$, this would give rise to $\mathcal{O}\left(\frac{1}{\epsilon}\right)$ terms and would modify the constant term, which reflects that these terms are scheme-dependent. In section \ref{sec:minvol}, we will obtain the same structure at strong coupling by computing the minimal volume of a 2-brane in AdS, which suggests that an anomalous dual conformal Ward identity fixes the finite part of the 4-point amplitude to all orders as it does in $\mathcal{N}=4$ SYM.

% ***************************************
\subsection{Bubbles and Triangles}\label{ssub:bubt}
\begin{figure}
\centering
\parbox{5cm}{
\includegraphics[width=4.8cm]{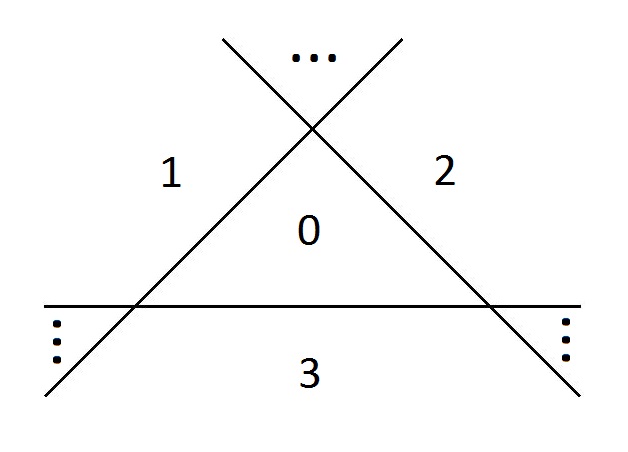}
\caption{Triangle diagrams}
\label{triangle}}
\qquad
\begin{minipage}{5cm}
\hspace*{0.9cm}\includegraphics[width=2.9cm]{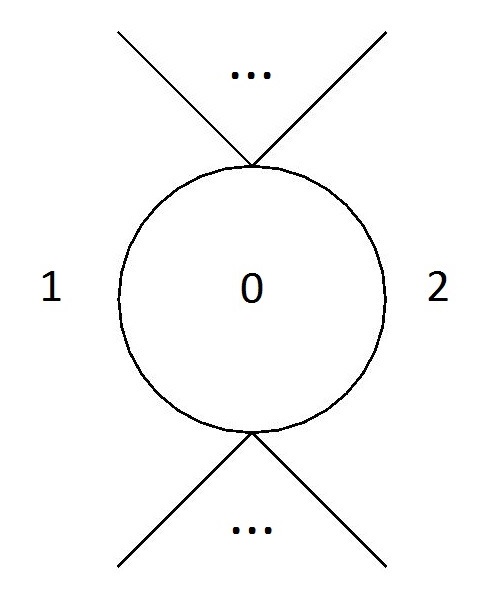}
\caption{Bubble diagrams}
\label{bubble}
\end{minipage}
\end{figure}

% ***************************************
In the previous section, we focused on 1-loop box diagrams. In principle, one can also consider triangle or bubble diagrams, as depicted in Figures \ref{triangle} and \ref{bubble}. Dual conformal symmetry restricts triangle integrals to have the form
  
\[
\mathcal{M}_{triangle}=\int d^{6}x_{0}\frac{\left(x_{12}^{2}\right)^{\alpha}\left(x_{23}^{2}\right)^{\beta}\left(x_{31}^{2}\right)^{\gamma}}{\left(x_{01}^{2}\right)^{\nu_{1}}\left(x_{02}^{2}\right)^{\nu_{2}}\left(x_{03}^{2}\right)^{\nu_{3}}}
\]
where $\nu_1+\nu_2+\nu_3=6$ and $\left(\alpha,\beta,\gamma\right)=\frac{1}{2}\left(\nu_{1}+\nu_{2}-\nu_{3},-\nu_{1}+\nu_{2}+\nu_{3},\nu_{1}-\nu_{2}+\nu_{3}\right)$. From this equation, we see that such contributions only exist for $n>4$ legs (for $n=4$, they collapse to bubble integrals which we will describe shortly).
Furthermore, using the results of \cite{Davydychev:1995mq}, one finds that they evaluate to rational functions
of the kinematic invariants:
\[
\mathcal{M}_{triangle}=\pi^{n/2}i^{1-n} \left(x_{12}^{2}\right)^{\alpha+\nu_{3}-3}\left(x_{23}^{2}\right)^{\beta+\nu_{1}-3}\left(x_{31}^{2}\right)^{\gamma+\nu_{2}-3}\Pi_{i=1}^{3}\frac{\Gamma\left(3-\nu_{i}\right)}{\Gamma\left(\nu_{i}\right)}.
\]

For any number of legs, one can define the following dual conformal
1-loop bubble integrals:
\[
\mathcal{M}_{bubble}=\int d^{6}x_{0}\left(\frac{x_{ij}^{2}}{x_{0i}^{2}x_{0j}^{2}}\right)^{3}.
\]
Note that this integral must be a function of the kinematic invariant $x_{ij}^2$ but has mass dimension zero. Dimensional 
analysis therefore implies that it must evaluate to a constant, which can be explicitly verified using the formulae in 
Appendix A of \cite{SmirnovBook}. Hence, a bubble contribution would modify the constant term of the 1-loop 4-point amplitude in \eqref{1loop4pt}, 
but as we explained in the previous section, the constant term is scheme-dependent in any case. 

%--------------------------------------------------
\subsection{Higher Loops} \label{2loop}
%--------------------------------------------------

\begin{figure}[h]
 \centering
 \includegraphics[width=0.39\textwidth]{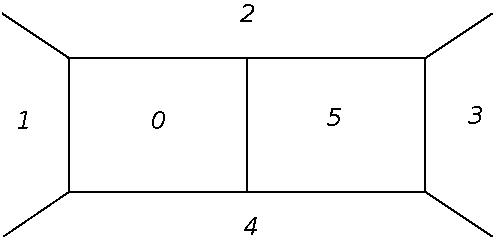}
 \caption{Two loop box.}
 \label{fig:2figsB}
\end{figure}

\begin{figure}
 \centering
 \includegraphics[width=0.8\textwidth]{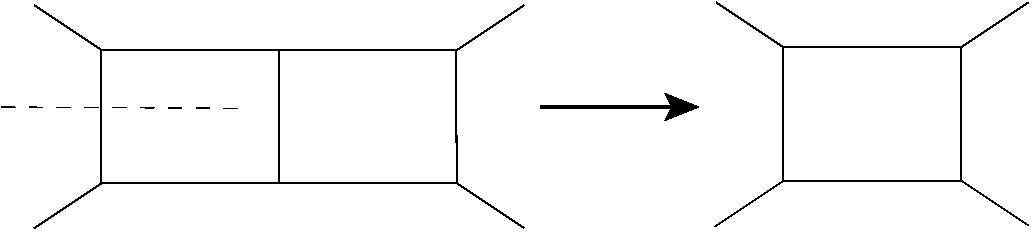}
 \caption{Cutting a propagator of the 2-loop box gives a 1-loop box.}
 \label{fig:cut}
\end{figure}

In this section we will consider 2-loop 4-point amplitudes with dual conformal symmetry in six dimensions. 
In $\mathcal{N}=4$ SYM, the 4-point 2-loop integrand has a double box topology, as depicted in Figure \ref{fig:2figsB}. 
To simplify the analysis we will focus on double box diagrams, although contributions with triangle or bubble subdiagrams deserve further study. In this case, there are ten possible structures consistent 
with dual conformal symmetry in $D=6$, two of which reduce to the 1-loop 
integrand in \eqref{4pt1loopintegrand} after 
cutting the propagators with momentum $x_{01}$ or $x_{53}$, as depicted in Figure \ref{fig:cut}. 
These two integrands are given by
\begin{equation}\label{2loopspl}
\begin{split}
& \int d^{D}x_{0}d^{D}x_{5}\frac{\left(x_{13}^{2}\right)^{2}\left(x_{24}^{2}\right)^{2}}{\left(x_{01}^{2}\right)^{2}x_{02}^{2}x_{04}^{2}x_{52}^{2}\left(x_{53}^{2}\right)^{2}x_{54}^{2}\left(x_{05}^{2}\right)^{2}}+s\leftrightarrow t
\\
& \int d^{D}x_{0}d^{D}x_{5}\frac{x_{13}^{2}\left(x_{24}^{2}\right)^{4}}{x_{01}^{2}\left(x_{02}^{2}\right)^{2}\left(x_{04}^{2}\right)^{2}\left(x_{52}^{2}\right)^{2}x_{53}^{2}\left(x_{54}^{2}\right)^{2}x_{05}^{2}}+s\leftrightarrow t .
\end{split}
\end{equation}
For example, if we cut the propagator with momentum $x_{01}$ (which amounts to taking this momentum on-shell), 
then the momenta $x_{02}$ and $x_{04}$ become on-shell and we can re-label $x_{05}$ as $x_{15}$ since $x_0$ is 
no longer integrated over. Discarding on-shell momenta, the two integrands above reduce to terms appearing in the 1-loop 
integrand \eqref{4pt1loopintegrand}. 

In appendix \ref{app:6d4p2l}, we list all ten integrands consistent 
with dual conformal symmetry as well as their Mellin-Barnes representations. 
The Mellin-Barnes representation of the two distinguished integrals in \eqref{2loopspl} are 
remarkably similar to that of $\mathcal{N}=4$ SYM, even though their spacetime 
representation looks completely different. This suggests that the 2-loop 4-point 
amplitude of a 6d dual conformal theory should be very closely related to that of $\mathcal{N}=4$ SYM, as we found at 1-loop. 
It would be very interesting to integrate these expressions in $D=6-2 \epsilon$ dimensions and see if they arise from exponentiating the 1-loop amplitude, which would suggest a BDS-like formula. 

%*********************************************************************
\section{Minimal volumes in AdS}\label{sec:minvol}
%*********************************************************************

At strong coupling, the duality between $\mathcal N=4$ super-Yang-Mills and string theory on $AdS_5 \times S_5$ was exploited 
in \cite{Alday:2007hr} to verify the validity of the BDS form of the four point amplitude. This was performed by finding an extremal surface $\Sigma$ in $AdS_5$, which ends on a four-sided null-polygon at the boundary. 
This boundary polygon can be thought of as the contour of a Wilson-loop dual to the four point amplitude.
The extremal surface $\Sigma$ then constitutes the world-sheet of the probe-string ending on the boundary Wilson-loop. 
The extremality of $\Sigma$ implies that the string world-sheet satisfies the equations of motion following from 
Nambu-Goto action. The four point amplitude is then obtained by evaluating the on-shell action for the string world-sheet and exponentiating.

In this section we will find a simple generalization of the Alday-Maldacena solution that minimizes the action for a 2-brane in $AdS_d$, with $d\geq 5$. Note that 11d supergravity admits a solution of $AdS_7\times S^4$ with the 3-form potential having non-zero legs only along the sphere, which describes a stack of M5-branes in the supergravity limit. Hence, our solution can then be interpreted as a single $M2$-brane in the near horizon geometry of a stack of M5-branes (since the solution is localized on the sphere, the 3-form can be neglected)\cite{BERGSHOEFF198775,Ortin:1997jh}.

% ++++++++++++++++++++++++++++++++++++++++++++++++++++++++
\subsection{General setup}
% ++++++++++++++++++++++++++++++++++++++++++++++++++++++++

The action for an M2 brane is given by the induced volume on the brane 
\begin{equation}\label{acgen}
\mathcal{S}=\int d^{3}\xi\mathcal{L}=\mathcal{T}_{M2}\int d^{3}\xi\sqrt{{\text{Det}}\left(G_{\mu\nu}\frac{\partial X^{\mu}}{\partial\xi^{a}}\frac{\partial X^{\nu}}{\partial\xi^{b}}\right)},
\end{equation}
where $\mathcal{T}_{M2}$ is the M2-brane tension, $\left(\xi_{1},\xi_{2},\xi_{3}\right)$ are the worldvolume coordinates of the brane, and $X^\mu$ are the coordinates of the target space with metric $G_{\mu \nu}$. We will take the target space to be $AdS_d$ (with $d \geq 5$). In the Poincare patch the metric is 
\begin{equation}\label{padsmet}
G_{\mu\nu}dX^{\mu}dX^{\nu} = \frac{-dy_0^2 + r^2 + \Sigma y_i^2}{r^2}.
\end{equation}
For the rest of the discussion we will set 
\begin{equation}
 y_i= 0, \forall ~i=4,5,\dots
\end{equation}
and focus on a $AdS_5$ slice of \eqref{padsmet}. This $AdS_5$ can be embedded in $\mathbb{R}^{2,4}$ by the relation 
\begin{equation}\label{poincond}
 - Y_{-1}^2  - Y_{0}^2 + Y_{1}^2 + Y_{2}^2 + Y_{3}^2 + Y_{4}^2  = -1.
\end{equation}
These coordinates are related to the Poincare coordinates by 
\begin{equation}\label{gloco}
 Y^{\mu} = \frac{y^\mu}{r}, ~\text{where}~ \mu = 0,1,2,3, ~~ Y_{-1} + Y_4 = \frac{1}{r}, ~~ Y_{-1} - Y_4 = \frac{r^2 + y^{\mu} y_{\mu}}{r}.
\end{equation}
% 

% ++++++++++++++++++++++++++++++++++++++++++++++++++++++++
\subsection{Single cusp solution}
% ++++++++++++++++++++++++++++++++++++++++++++++++++++++++

In terms of embedding coordinates, the single cusp solution for a 2-brane satisfies the following constraints:
\begin{equation}\label{glocusp}
 Y_3 = 0,~~Y_0^2 - Y_{-1}^2 = Y_1^2 - Y_4^2.
\end{equation}
This is essentially the same as the single-cusp Alday-Maldacena solution with one constraint ($Y_2=0$) removed, which can be motivated by noting that a $2$-brane is one dimension higher than a string and therefore 
requires one less condition to specify it. Nevertheless, it is rather nontrivial that the constraints in \eqref{glocusp} actually describe a 2-brane solution. Indeed, when the constraints are transformed to Poincare coordinates using \eqref{gloco}, they have a very different structure than the Alday-Maldacena solution:
\begin{equation}
\begin{split}\label{cuspconfigexp}
 r = \sqrt{2 \left( y_0^2 - y_1^2 \right) - y_2^2},  ~~y_3=0.
\end{split}
\end{equation}
If we use the worldvolume diffeomorphsim symmetry to set $\left(\xi_{1},\xi_{2},\xi_{3}\right)=\left(y_{0},y_{1},y_{2}\right)$, 
we find that \eqref{cuspconfigexp} is indeed a solution to the Nambu-Goto equations of motion for a 2-brane in AdS. 
After realizing how to lift the single cusp solution for a string to that of a 2-brane, it is straightforward to generate a four-cusp solution for a 2-brane following the procedure in \cite{Alday:2007hr}, which we shall describe in the next section. 

% ++++++++++++++++++++++++++++++++++++++++++++++++++++++++
\subsection{Four cusp solution} \label{ssec:minvol4cusp}
% ++++++++++++++++++++++++++++++++++++++++++++++++++++++++

On performing a set of $SO(2,4)$ transformations following \cite{Alday:2007hr}, 
the constraints in \eqref{glocusp} become 
\begin{equation}\label{seqtconfigglo}
 Y_4 = 0 , ~~ Y_0 Y_{-1} = Y_1 Y_2. 
\end{equation}
In terms of Poincare coordinates \eqref{seqtconfigglo} translates to 
\begin{equation}\label{seqtconfig}
 y_0 = y_1 y_2,~~r = \sqrt{\left(1 - y_1^2 \right)\left(1 - y_2^2 \right)- y_3^2}, 
\end{equation}
where we have made use of \eqref{gloco}. Choosing $\left(\xi_{1},\xi_{2},\xi_{3}\right)=\left(y_{0},y_{1},y_{2}\right)$ gives a 4-cusp solution to the equations of motion following from \eqref{acgen}. Note that this solution describes a minimal-volume 2-brane in AdS and can be obtained from the minimal-area string solution in \cite{Alday:2007hr} by writing it in homogenous coordinates and relaxing a constraint ($Y_3=0$). In fact, it can be generalized to an infinite family of $p$-brane solutions in AdS:
\[
 y_0 = y_1 y_2,~~r = \sqrt{\left(1 - y_1^2 \right)\left(1 - y_2^2 \right)-\sum_{i=3}^{p+1} y_i^2}.
\]
These solutions have the remarkable property that their on-shell actions factorize into the on-shell action of the minimal string solution times a contribution from the remaining $p-2$ directions of the world-volume, although this factorization does not occur at the level of the classical solutions themselves. 

\begin{figure}[htb]
\centering
  \begin{tabular}{@{}ccc@{}}
    \includegraphics[width=.45\textwidth]{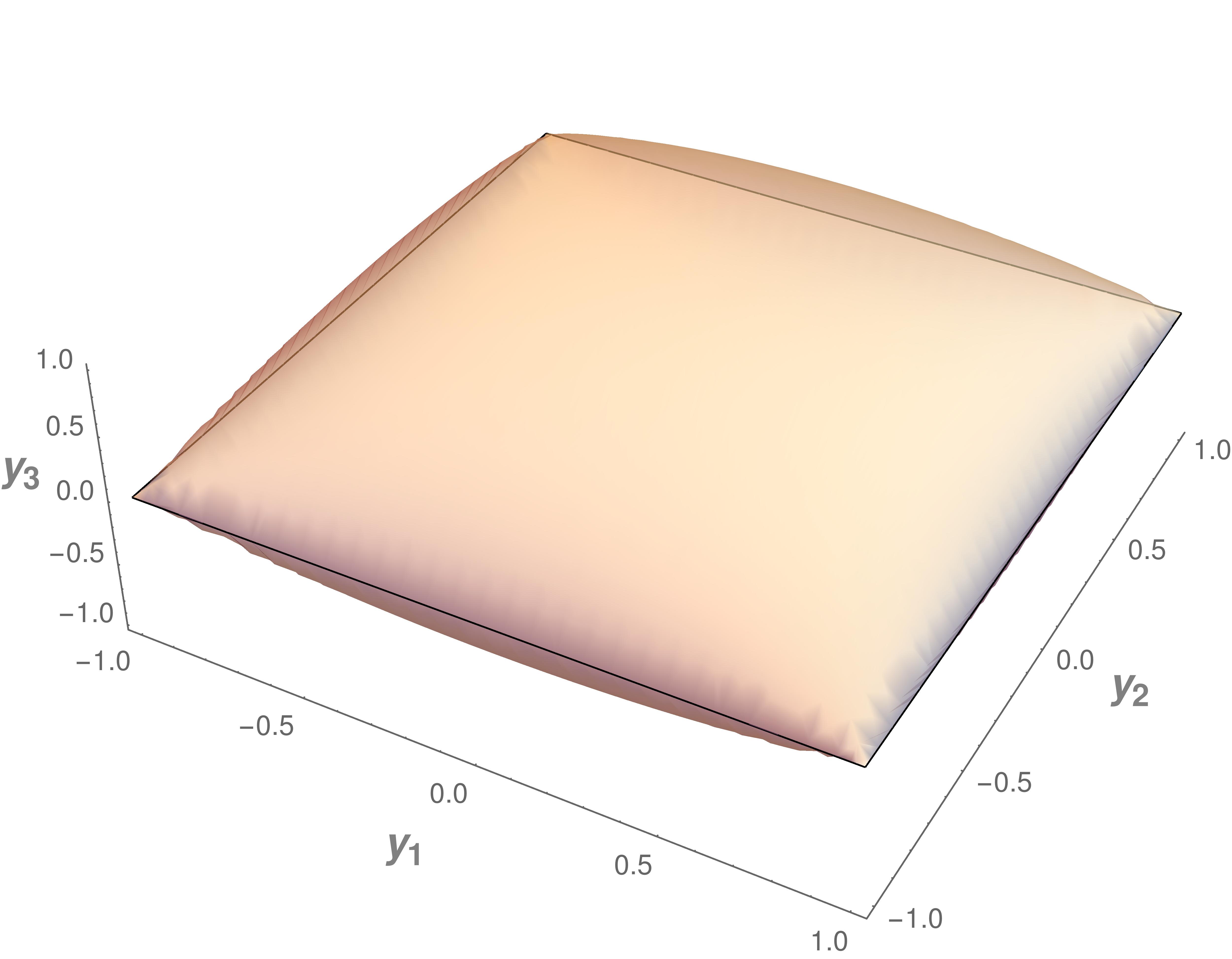} &
    \includegraphics[width=.45\textwidth]{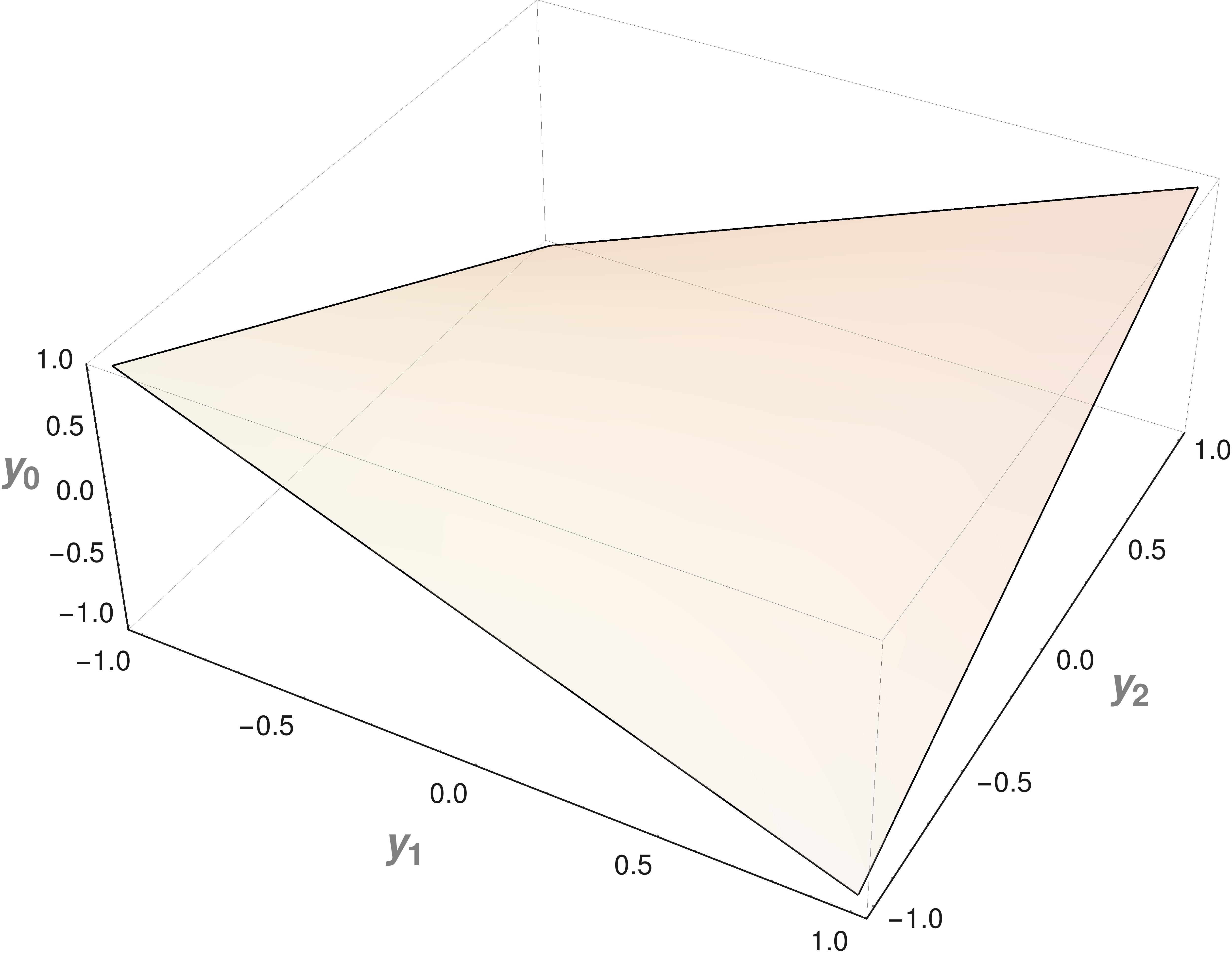} \\
  \end{tabular}
  \caption{Plots of the boundary surface \eqref{bdyembed}, obtained by projecting it on the space orthogonal to 
$y_0$, and $y_3$ respectively, for $b=0$ $(s=t)$. When projected orthogonal to $y_3$, the surface reduces to a null polygon with its interior filled.}\label{fig:bdysur}
\end{figure}

Note that the $r \rightarrow 0$ limit of this solution is a two dimensional surface on the boundary of 
$AdS$ which may be taken as the contour of a Wilson surface in the 6d dual field theory. In fig. \ref{fig:bdysur}, we have plotted this surface by projecting it on the spaces orthogonal to 
$y_0$ and $y_3$, respectively. 
From this figure, we see that the Wilson surface can be visualized as 
a pillow with four seams corresponding to the edges of a null polygon. 
Moreover, if we identify the edges with the momenta of a scattering amplitude, the resulting Mandelstam variables satisfy $s=t$. 

To generalize this solution to $s \neq t$ we can simply apply a boost to \eqref{seqtconfigglo} followed by a rescaling following \cite{Alday:2007hr} to obtain
\begin{equation}\label{bofulsol}
\begin{split}
& y_0 = \frac{a \sqrt{1+b^2} ~\sinh{\xi_1}~\sinh{\xi_2} }{\cosh{\xi_1}~\cosh{\xi_2} + b ~\sinh{\xi_1}~\sinh{\xi_2} },
 ~~y_1 = \frac{a  ~\sinh{\xi_1}~\cosh{\xi_2} }{\cosh{\xi_1}~\cosh{\xi_2} + b ~\sinh{\xi_1}~\sinh{\xi_2} },\\
&y_2 = \frac{a  ~\cosh{\xi_1}~\sinh{\xi_2} }{\cosh{\xi_1}~\cosh{\xi_2} + b ~\sinh{\xi_1}~\sinh{\xi_2} }, 
~~y_3 = \frac{a ~\xi_3 }{\cosh{\xi_1}~\cosh{\xi_2} + b ~\sinh{\xi_1}~\sinh{\xi_2} },\\
&\qquad \qquad \qquad \qquad~~~~~~~
r = \frac{a ~\sqrt{1-\xi_3^2}}{\cosh{\xi_1}~\cosh{\xi_2} + b ~\sinh{\xi_1}~\sinh{\xi_2} },
 \end{split}
\end{equation}
where the world sheet coordinates $\{\xi_1,\xi_2,\xi_3\}$, must take values in $[-1,1]$ and we make the following 
identification of the Mandelstam variables with the parameters $a,~b$ of the solution:
\begin{equation}
 -s = \frac{8 a^2}{(2 \pi)^2 (1-b)^2}, ~-t = \frac{8 a^2}{(2 \pi)^2 (1+b)^2}.
\end{equation} 
It is not difficult to check that \eqref{bofulsol} solves the equations 
of motion arising from \eqref{acgen} and reduces to \eqref{seqtconfig} when $b=0$, which corresponds to $s=t$.  Also note that 
if we replace $\xi_3$ by an arbitrary function of $f(\xi_3) \in [-1,1]$ in \eqref{bofulsol}, it remains a valid solution 
of the equations of motion. This is simply related to 
world-sheet diffeomorphism invariance. 

Taking $r\rightarrow 0$ gives the following equations for the surface in boundary:
\begin{equation}
 \begin{split}\label{bdyembed}
& y_0 = \frac{a \sqrt{1+b^2} ~\sinh{\xi_1}~\sinh{\xi_2} }{\cosh{\xi_1}~\cosh{\xi_2} + b ~\sinh{\xi_1}~\sinh{\xi_2} },
 ~~y_1 = \frac{a  ~\sinh{\xi_1}~\cosh{\xi_2} }{\cosh{\xi_1}~\cosh{\xi_2} + b ~\sinh{\xi_1}~\sinh{\xi_2} },\\
&y_2 = \frac{a  ~\cosh{\xi_1}~\sinh{\xi_2} }{\cosh{\xi_1}~\cosh{\xi_2} + b ~\sinh{\xi_1}~\sinh{\xi_2} }, 
~~y_3 = \frac{a}{\cosh{\xi_1}~\cosh{\xi_2} + b ~\sinh{\xi_1}~\sinh{\xi_2} }.\\
 \end{split}
\end{equation} 
Although these equations are identical to the 4-cusp string solution of Alday-Maldacena solution \cite{Alday:2007hr}, they are now in Minkowski space rather than AdS, which 
makes the two surfaces distinct.

On plugging the solution \eqref{bofulsol} back into Lagrangian \eqref{acgen} and restoring the AdS radius $R$, we get 
\begin{equation}
 \mathcal L = \frac{R^3 \mathcal{T}_{M2}}{(1-\xi_3^2)^2},
\end{equation}
which is independent of the coordinates $\xi_1$ and $\xi_2$. This is not surprising, since $\xi_1, ~\xi_2$ 
are the corresponding coordinates of the string world-sheet in \cite{Alday:2007hr}, where 
the Lagrangian simply evaluated to unity on substituting the full solution. In the next section, we will describe how to evaluate the on-shell action, which is divergent and therefore needs to be regulated.

% ++++++++++++++++++++++++++++++++++++++++++++++++++++++++
\subsection{Regulated on-shell action} \label{ssec:regac}
% ++++++++++++++++++++++++++++++++++++++++++++++++++++++++
To implement the analogue of dimensional regularization in the bulk, we follow the 
method outlined in \cite{Alday:2007hr} and define the regularized action as 
\begin{equation}\label{regacsch}
 \mathcal S_{\text{reg}} = \int \frac{\mathcal L_{\epsilon = 0}}{r^\epsilon},
\end{equation}
where $\epsilon<0$. In \cite{Alday:2007hr}, this regularization was motivated by analytically continuing the near-horizon metric of black D-banes. Since we are computing the minimal volume of an M2-brane in the bulk, it would be interesting to first dimensionally reduce AdS$_7$ on a circle and then apply the above procedure to compute the on-shell action of the dimensionally reduced solution. In this paper, we will stick with regularization in \eqref{regacsch} for simplicity.   

Although it is difficult to obtain a 4-cusp solution for the regularized action in \eqref{regacsch}, it suffices to know the single-cusp solution for the regularized action. In particular, the single-cusp solution in \eqref{cuspconfigexp} generalizes to 
\begin{equation}\label{regcuspsol}
 r(y_0,y_1,y_2) = \sqrt{(2+\epsilon) \left( y_0^2 - y_1^2 \right) - y_2^2},  ~~y_3(y_0,y_1,y_2)=0,
\end{equation}
as a solution of the regularized action \eqref{regacsch}.
Now we wish to modify the solution in \eqref{bofulsol}, so that it reduces to the regulated solution \eqref{regcuspsol}
near the cusps. This can be accomplished by the following redefinitions:
\begin{equation}\label{regfulsol}
 r \rightarrow \sqrt{1+\frac{\epsilon}{2}} ~r_{\epsilon=0}, ~y_3 \rightarrow \sqrt{1+\frac{\epsilon}{2}} ~(y_3)_{\epsilon=0}, 
~ y_i \rightarrow {(y_i)}_{\epsilon=0}, ~\text{for} ~i=0,1,2.
\end{equation}
where the $\epsilon=0$ solution refers to that in \eqref{bofulsol}.

Upon substituting the solution in \eqref{regfulsol} into the regulated action \eqref{regacsch}, we find that the integral over $\xi_3$ decouples from the $\xi_1,\xi_2$ integrals and we obtain
\begin{equation}
 i \mathcal S_{\text{reg}} =\mathcal {T}_\epsilon \tilde{\mathcal{S}} 
\label{osat}
\end{equation}
where 
\[
\mathcal{T}_{\epsilon} =R^3 \mathcal{T}_{M2} \int \frac{d\xi_3}{\left(1-\xi_3^2\right)^{2+\frac{\epsilon}{2}}}=-\frac{R^{3}\mathcal T_{M_{2}}}{\epsilon}+\mathcal{O}\left(\epsilon^{0}\right)
\]
and \footnote{Note that  $\tilde{\mathcal{S}}$ is identical to the regulated on-shell action for the string solution obtained in \cite{Alday:2007hr} apart from an additive constant. In \cite{Alday:2007hr}, this constant was reported to be $1/2$ instead of $3/2$. For more details, see Appendix \ref{app:caldet}.}
\[
\tilde{\mathcal{S}}=-\frac{\pi\Gamma\left(-\frac{\epsilon}{2}\right)^{2}}{\Gamma\left(\frac{1-\epsilon}{2}\right)^{2}}\left[_{2}F_{1}\left(\frac{1}{2},-\frac{\epsilon}{2},\frac{1-\epsilon}{2};b^{2}\right)+\frac{3}{2}\right]+\mathcal{O}\left(\epsilon\right).
\]
Recalling that $\mathcal T_{M_{2}}=\frac{1}{\left(2\pi\right)^{2}l_{p}^{3}}$ and that in the near-horizon limit of a stack of $N$ M5-branes $\left(\frac{R}{l_{p}}\right)^{3}=\pi N $, we finally obtain
\begin{equation}
i \mathcal S_{\text{reg}}=-\frac{N}{4\pi\epsilon}\left[2 \mathcal S_{div,s}+2 \mathcal S_{div,t}+\frac{1}{4}\ln^{2}\left(\frac{s}{t}\right)+C\right],
\label{osaction}
\end{equation}
where $\mu$ is the renormalization scale,
\[
\mathcal S_{div,s}=-\frac{1}{\epsilon^{2}}\left(\frac{\mu^{2}}{s}\right)^{\epsilon/2}+\frac{2\ln2}{\epsilon}\left(\frac{\mu^{2}}{s}\right)^{\epsilon/2}
\]
(with a similar formula for $\mathcal S_{div,t}$), and 
\[
C=\frac{\pi^{2}}{3}-8\ln^{2}2-\frac{3}{2}.
\]
Note that the $\mathcal{O}\left(\frac{1}{\epsilon}\right)$ terms in $\mathcal{S}_{div}$ and the constant term $C$ are scheme-dependent since 
they can be altered by rescaling $\mu$. 

In terms of the on-shell action for the minimal 2-brane in \eqref{osaction}, the expectation value of the Wilson surface with contour in \eqref{bdyembed} is given by 
\[
 \mathcal M_4 = e^{i \mathcal S_{\text{reg}}}.
\]
Note that the on-shell action has a very similar structure to the 1-loop 4-point amplitude in \eqref{1loop4pt}. This suggests that the 6d 4-point amplitude is dual to a Wilson surface and can be described to all-loop orders in terms of a BDS-like formula. A novel feature of the on-shell action for the 2-brane compared to the on-shell action of the string obtained in \cite{Alday:2007hr} is an overall prefactor which diverges as $\epsilon \rightarrow 0$. A similar divergence was found in previous holographic calculations of Wilson surfaces of the 6d $(2,0)$ theory \cite{Berenstein:1998ij} and circular Wilson loops of 5d SYM \cite{Young:2011aa}, and can be associated with the conformal anomaly of Wilson surfaces \cite{Graham:1999pm,Henningson:1999xi,Gustavsson:2003hn,Gustavsson:2004gj}. Furthermore, it may be possible to interpret the prefactor in \eqref{osaction} which goes as $N/\epsilon$ as an  t'Hooft coupling in the $6-2\epsilon$ dimensions. We shall
discuss this point further in \S \ref{sec:disco}.

% **********************************************************
\section{Discussion}\label{sec:disco}
% **********************************************************

In this note, we have explored dual conformal symmetry for loop 
amplitudes in six dimensions. This study is motivated by the fact that the scattering amplitudes of D3 and M2-brane worldvolume theories have dual conformal symmetry 
even though it is hidden from the point of view of their Lagrangian formulations.
This suggests that it may also play a role in the scattering amplitudes of 
M5-brane worldvolume theories. 

If we consider a hypothetical 6d theory which has a planar limit 
and scattering amplitudes with rational (field theoretic) loop integrands, we find that imposing dual 
conformal symmetry fixes the integrand for the 
4-point 1-loop amplitude. After performing the integral over 
loop momentum in $d=6-2\epsilon$ dimensions, we obtain a result that is 
remarkably similar to the 4-point amplitude of $\mathcal{N}=4$ SYM and the ABJM theory. 
We also analyze two-loop 4-point amplitudes in six dimensions and identify two integrands compatible with dual 
conformal symmetry and unitarity. 

Given that the planar scattering amplitudes in $\mathcal{N}=4$ SYM can be computed at strong from the area of a string worldsheet in AdS whose boundary corresponds to a null-polygonal 
Wilson loop, one may expect that the amplitudes of the 6d theory can also be computed holographically by minimizing 
the volume of a 2-brane whose boundary corresponds to a Wilson surface which somehow 
encodes a null polygon. In fact, we find a simple generalization of the 4-cusp 
Alday-Maldacena solution which minimizes the  action of a 2-brane in AdS and verify that 
its on-shell action has essentially the same structure as the 1-loop 4-point amplitude we deduced 
using dual conformal symmetry. This suggests that the 6d 4-point amplitude is dual to a Wilson 
surface and can be described to all-loop orders using a formula analogous to the BDS formula 
of $\mathcal{N}=4$ SYM. A new feature of the on-shell action for the 2-brane compared to 
that of the string in AdS is an overall prefactor which diverges as $\epsilon \rightarrow 0$. 
This additional divergence was observed in previous holographic calculations of Wilson surfaces. 

Although our calculations are highly suggestive, the existence of a 6d theory with 
these properties and its relation to M5 branes is still rather speculative. 
In particular, there are a number of questions that deserve further study:

\begin{itemize}

\item
Perhaps the most immediate task would be to extend our perturbative 4-point calculations  
to higher loops and check if they are consistent with an all-loop formula 
for the 4-point amplitude. This would require extending our 1-loop calculation 
to $\mathcal{O}\left( \epsilon^2  \right)$, computing the 2-loop Mellin-Barnes integrals we obtained in 
appendix \ref{app:6d4p2l}, and determining if any linear combination of these integrals can arise from 
exponentiating the 1-loop amplitude. In the end, it may only be necessary to consider 
the two integrands we describe in section \ref{2loop}, whose Mellin-Barnes 
representations are very similar to that of the 2-loop 4-point amplitude of $\mathcal{N}=4$ SYM.
\item
Note that two of the propagators in the 1-loop integrand in 
\eqref{4pt1loopintegrand} are squared, which suggests that if there is an 
underlying Lagrangian, it contains terms of the form $\phi \partial^4 \phi$. 
Such theories are not generally unitary unless they arise from expansions of non-local 
theories like string theory \cite{Simon:1990ic}. This is certainly a possibility for the M5-brane worldvolume theory given that it has been argued to be a non-gravitational self-dual string theory \cite{Witten:1995em,Strominger:1995ac}. Another possibility is that the 6d theory in question is not unitary. Indeed, four-derivative terms appear in certain non-unitary 6d superconformal theories \cite{Ivanov:2005qf,Beccaria:2015uta}, so studying the scattering amplitudes of these theories is an important direction for future research. In \cite{Huang:2010rn}, 
it was proven that it is not possible to construct tree-level amplitudes of 
self-dual tensor multiplets assuming unitarity and locality. It would be interesting 
to revisit this question in light of our loop-level calculations, which suggest that 
some of these assumptions may need to be relaxed. 
\item A crucial assumption in our analysis is the existence
of a planar limit. In order to define such a limit, there must be a
coupling which can be sent to zero as the rank of the gauge group $N$ is 
sent to infinity. For the M2-brane worldvolume theory, this can be accomplished 
by performing a $\mathbb{Z}_k$ orbifold of the space transverse to the M2-branes, 
which ultimately gives rise to the 't Hooft coupling $N/k$. 
For the M5-brane worldvolume theory, our results suggest that the coupling is tied 
to the dimension of the worldvolume itself. 
In particular, note that the prefactor of the on-shell action in \eqref{osaction} is 
proportional to $N/\epsilon$. If we interpret this as the 't Hooft coupling, 
we see that the coupling diverges as $d\rightarrow 6$. 
This is reminiscent of the conjecture that 5d SYM is equivalent to the 6d $(2,0)$ 
compactified on a circle \cite{Douglas:2010iu,Lambert:2010iw}, 
where the radius of compactification is proportional to the 5d 
coupling \cite{Witten:2009at}. It would therefore be interesting to 
explore how our results relate to the planar scattering amplitudes and 
Wilson loops of 5d SYM. Although perturbative amplitudes of 5d SYM have 
UV divergences \cite{Bern:2012di}, it has been argued that they may 
be removed by non-perturbative effects \cite{Papageorgakis:2014dma}. 
\item 
In $\mathcal{N}=4$ SYM, dual superconformal 
symmetry was shown to arise from self-duality of the dual string theory under a combination of bosonic and fermionic T-duality transformations. It would therefore be interesting to look for an analogue of T-duality in M-theory, 
perhaps by compactifying the theory on a circle to obtain IIA string theory, 
performing various combinations of bosonic and fermionic T-duality transformations, 
and then decompactifying back to M-theory. Some of these ideas 
were considered in \cite{Jeon:2012fn}.
\end{itemize}
Ultimately, we hope this line of investigation will yield new insight into the 
longstanding question of how to formulate the worldvolume theory of M5-branes.

% %~~~~~~~~~~~~~~~~~~~~~~~~~~~~~~~~~~~~~~~~~~~~~~
\acknowledgments 
% %~~~~~~~~~~~~~~~~~~~~~~~~~~~~~~~~~~~~~~~~~~~~~~

We thank Paul Heslop, Simon Ross, Douglas Smith, and Tadashi Takayanagi for 
helpful comments on the manuscript. JB is supported by the STFC Consolidated Grant ST/L000407/1. 
AL is supported by the Royal Society as a Royal Society University Research Fellowship holder.

%%~~~~~~~~~~~~~~~~~~~~~~~~~~~~~~~~~~~~~~~~~~~~~~
\appendix
%%~~~~~~~~~~~~~~~~~~~~~~~~~~~~~~~~~~~~~~~~~~~~~~

% ********************************************************************************
\section{Details of 2-loop 4-point amplitude}\label{app:6d4p2l}
% ********************************************************************************

In general, the 2-loop integrand with a double box topology has the following form:
\begin{equation}\label{2loops}
\mathcal{A}_{4}^{(2)}=
\int d^{6-2\epsilon}x_{0}d^{6-2\epsilon}x_{5}
\frac{\left(x_{13}^{2}\right)^{\alpha_{1}}\left(x_{24}^{2}\right)^{\alpha_{2}}}
{\left(x_{01}^{2}\right)^{\nu_{2}}\left(x_{02}^{2}\right)^{\nu_{1}}\left(x_{04}^{2}\right)^{\nu_{3}}
\left(x_{52}^{2}\right)^{\nu_{6}}\left(x_{53}^{2}\right)^{\nu_{5}}\left(x_{54}^{2}\right)^{\nu_{4}}
\left(x_{05}^{2}\right)^{\nu_{7}}}+s\leftrightarrow t.
\end{equation}
There are ten possibilities consistent with dual conformal symmetry, which we summarize in table \ref{tab:2loopexp}.
\begin{table}
\centering
\caption{Table listing sets of exponents in \eqref{2loops} consistent with dual conformal symmetry.}\label{tab:2loopexp}
\begin{tabular}{|c|c|c|c|c|c|c|c|c|c|}
\hline
No. & $\nu_1$ & $\nu_2$ & $\nu_3$ & $\nu_4$ & $\nu_5$ & $\nu_6$ & $\nu_7$ & $\alpha_1$ & $\alpha_2$ \\
\hline
\hline
(1)& 1  & 2  & 1  & 1  & 2  & 1  & 2  & 2  & 2  \\
(2)&2  & 1  & 2  & 2  & 1  & 2  & 1  & 1  & 4  \\
(3)&1  & 1  & 1  & 1  & 1  & 1  & 3  & 1  & 2  \\
(4)&2  & 1  & 1  & 2  & 1  & 1  & 2  & 1  & 3  \\
(5)&1  & 1  & 2  & 1  & 1  & 2  & 2  & 1  & 3  \\
(6)&3  & 1  & 1  & 3  & 1  & 1  & 1  & 1  & 4  \\
(7)&1  & 1  & 3  & 1  & 1  & 3  & 1  & 1  & 4  \\
(8)&1  & 2  & 2  & 1  & 2  & 2  & 1  & 2  & 3  \\
(9)&2  & 2  & 1  & 2  & 2  & 1  & 1  & 2  & 3  \\
(10)&1  & 3  & 1  & 1  & 3  & 1  & 1  & 3  & 2  \\
\hline
\end{tabular}
\end{table}
The integral in \eqref{2loops} has a Mellin-Barnes representation
\begin{equation}
 \begin{split}
 & \mathcal I(\nu_1,\nu_2,\nu_3,\nu_4,\nu_5,\nu_6,\nu_7,\alpha_1, \alpha_2) \\
 & = s^{\alpha_1+D-\nu} t^{\alpha_2} \frac{(-1)^{D+1} \pi^D}{ \Gamma(D-\nu_{4567})\prod_{l=2,4,5,6,7} \Gamma(\nu_l) } 
 \int \prod_{j=1}^4 \frac{dz_j}{2 \pi i} \left(\frac{t}{s}\right)^{z_1} \Gamma(\nu_2 + z_1) \Gamma(-z_1) \\
 & \times \frac{\Gamma(z_2 + z_4)\Gamma(z_3 + z_4)\Gamma(\nu_{123} -D/2 + z_4)\Gamma(\nu_7 + z_1 -z_4)}
 {\Gamma(\nu_1 z_3 +z_4 )\Gamma(\nu_3 + z_2 + z_4)\Gamma(D-\nu_{123} +z_1-z_4)} \Gamma(-z_2 -z_3-z_4)\\
 & \times \Gamma( \nu_5 +z_1 +z_2 +z_3 +z_4) \Gamma(\nu_{4567} -D/2 +z_1 -z_4 )\Gamma( D/2 -\nu_{12} +z_2 )\\
 & \times \Gamma(D/2 - \nu{23} + z_3 ) \Gamma(D/2 - \nu_{567} -z_1 -z_2)\Gamma(D/2- \nu{457} - z_1 - z_3), 
 \end{split}
\end{equation}
where $D = 6-2 \epsilon$, $\nu = \sum_{i=1}^7 \nu_i$ , $\nu_{ij} = \nu _i + \nu _j$, and $\nu_{ijk} = \nu_{i}+ \nu_j + \nu_k$.
Note that first and the second sets of exponents listed in table \ref{tab:2loopexp} correspond to the 
integrals \eqref{2loopspl}, and we shall denote their Mellin Barnes integrands as $I^{6d}_{(1)}$ and $I^{6d}_{(2)}$, respectively. 
For comparison, the Mellin-Barnes representation of the planar 
4-point 2-loop amplitude in $\mathcal{N}=4$ SYM is given by $\mathcal I(1,1,1,1,1,1,1,1,0)$ and we will denote the Mellin-Barnes integrand as $I^{4d}$. Remarkably, the Mellin-Barnes integrands corresponding to \eqref{2loopspl} are very similar to that of $\mathcal N=4$ SYM:
\begin{equation}
\begin{split}
 & \frac{I^{4d}(z_1,z_2,z_3,z_4)}{I^{6d}_{(1)}(z_1-1,z_2,z_3,z_4)} =  - \frac{t \Gamma(-z_1)}{\pi^2 \Gamma(1-z_1)} ,\\
 & \frac{I^{4d} (z_1,z_2,z_3,z_4)}{I^{6d}_{(2)}(z_1,z_2,z_3,z_4)} = 
 \frac{s^2 \Gamma (z_2+z_4+2) \Gamma (z_3+z_4+2) \Gamma (z_4+\epsilon +1) \Gamma
   (z_1-z_4+\epsilon +2)}{\pi ^2 \Gamma (z_2+z_4+1) \Gamma (z_3+z_4+1) \Gamma
   (z_4+\epsilon +2) \Gamma (z_1-z_4+\epsilon +3)}.
 \end{split}
\end{equation}

% ***********************************************************************************************************
\section{Details of regularized action}\label{app:caldet}
% ***********************************************************************************************************
 The regulated on-shell action in \eqref{osat} takes the form
\footnote{Note that on evaluating the onshell Nambu-Goto action, in the $\mathcal{O}(\epsilon)^2$ term, 
we find an extra additive $\frac{1}{4}$, compared to \cite{Alday:2007hr}. This eventually contributes to the constant $\mathcal{O}(\epsilon)^0$
term, in the amplitude. We think this should be present for the case of the string worldsheet in \cite{Alday:2007hr} as well.}
\begin{equation}
 i S = - T_\epsilon \int_{- \infty}^{+\infty} d\xi_1 d\xi_2  
 \frac{1+ \epsilon I_1 + \epsilon^2 \left( I_2 - 2 I_2^2 + \frac{1}{4} \right) }{\left( \cosh \xi_1 \cosh \xi_2 + b \sinh \xi_1 \sinh \xi_2\right)^{\epsilon}}
\end{equation}
where 
\begin{equation}
 \begin{split}
  I_1 &= \frac{\left(b^2-1\right) \cosh (2 \xi_1)+\left(b^2-1\right) \cosh (2 \xi_2)-2 \left(b^2+1\right)}{8 (b \sinh (\xi_1)
   \sinh (\xi_2)+\cosh (\xi_1) \cosh (\xi_2))^2}\\
  I_2 &=  \frac{-\left(b^2+1\right) \cosh (2 \xi_1) \cosh (2 \xi_2)-2 b \sinh (2 \xi_1) \sinh (2 \xi_2)+1+b^2}{16 (b \sinh
  (\xi_1) \sinh (\xi_2)+\cosh (\xi_1) \cosh (\xi_2))^2}\\
%   T &= \frac{\sqrt{\lambda_D C_D}}{2 \pi a^\epsilon} 
%   = \frac{ \left( \lambda (2 \pi)^{2 \epsilon} e^{\epsilon \gamma_E} \Gamma(2+\epsilon) \mu^{2\epsilon} \right)^{\frac{1}{2}}}{2 \pi a^\epsilon}
\mathcal  T_{\epsilon} &=R^3  \mathcal T_{M2} \int \frac{d\xi_3}{\left(1-\xi_3^2\right)^{2+\frac{\epsilon}{2}}} = 
R^{3}\mathcal T_{M_{2}}\frac{\sqrt{\pi}
\Gamma\left(-1-\frac{\epsilon}{2}\right)}{\Gamma\left(-\frac{1}{2}-\frac{\epsilon}{2}\right)}.
 \end{split}
\end{equation}
The integrals evaluate to 
% % \footnote{Note that the extra $\frac{1}{4}$ in the last integral, compared to \cite{Alday:2007hr}, contributes $-\frac{1}{\epsilon^2}+\dots$ to the 
% % final integral. {\JB{We might want to eventually kill this note.}}}
\begin{equation}
\begin{split}
 \int_{- \infty}^{+\infty} d\xi_1 d\xi_2   \frac{1}{\left( \cosh \xi_1 \cosh \xi_2 + b \sinh \xi_1 \sinh \xi_2\right)^{\epsilon}}
&= \frac{\pi  \Gamma \left(-\frac{\epsilon }{2}\right)^2}{\Gamma \left(\frac{1-\epsilon }{2}\right)^2}
 \, _2F_1\left(\frac{1}{2},-\frac{\epsilon }{2};\frac{1}{2}-\frac{\epsilon}{2};b^2\right), \\
  \int_{- \infty}^{+\infty} d\xi_1 d\xi_2   \frac{I_1}{\left( \cosh \xi_1 \cosh \xi_2 + b \sinh \xi_1 \sinh \xi_2\right)^{\epsilon}}
 &= \frac{2}{\epsilon} + \mathcal{O}\left(\epsilon^0\right),\\
 \int_{-\infty}^{+\infty} d\xi_1 d\xi_2   \frac{I_2 - 2 I_2^2+\frac{1}{4}}{\left( \cosh \xi_1 \cosh \xi_2 + b \sinh \xi_1 \sinh \xi_2\right)^{\epsilon}}
 &= -\frac{1}{2 \epsilon^2} + \mathcal{O}\left(\frac{1}{\epsilon}\right).
\end{split}
\end{equation}
Using the following expansions
\begin{equation}
 \begin{split}
 & \frac{\pi  \Gamma \left(-\frac{\epsilon }{2}\right)^2}{\Gamma \left(\frac{1-\epsilon }{2}\right)^2} = 
  \frac{4}{\epsilon ^2}-\frac{4 \log (4)}{\epsilon }+\left(2 \log ^2(4)-\frac{\pi ^2}{3}\right)+\mathcal O \left(\epsilon\right) \\
 &  _2F_1\left(\frac{1}{2},-\frac{\epsilon }{2};\frac{1}{2}-\frac{\epsilon}{2};b^2\right) = 
 \left( \frac{1}{2} \left( 1- b\right)^\epsilon + \frac{1}{2} \left( 1- b\right)^\epsilon \right) 
 - \frac{\epsilon^2}{4} \ln \left( \frac{1+b}{1-b}\right) + \mathcal O \left(\epsilon^3\right)\\
% %  & T = \frac{1}{2\pi} \sqrt{\frac{\lambda (2\pi)^{2\epsilon} \mu^{2 \epsilon} }{(8 a)^{2 \epsilon}}} 
% %  \left(1+\epsilon  \left(\frac{1}{2}+\log (8)\right)+\frac{1}{24} \epsilon ^2 \left(\pi ^2+3 (4 \log (8) (1+\log
% %    (8))-1)\right)+O\left(\epsilon ^3\right)\right) 
 & T_\epsilon = -\frac{R^3 \mathcal T_{M2}}{\epsilon} + \mathcal O \left( \epsilon^0 \right)
 \end{split}
\end{equation}
and putting all the terms together gives \eqref{osaction}.

%%%%%%%%%%%%%%%%%%%%%%%%%%%%%
\bibliographystyle{JHEP}
\bibliography{6dAmpM2}
\end{document}